# Cloud-Edge Collaborative Data Anomaly Detection in Industrial Sensor Networks

Tao Yang[1], Xuefeng Jiang[5], Wei Li[5], Peiyu Liu[2], Jinming Wang[3], Weijie Hao[4], Qiang Yang[5*]

[1]China Tobacco Zhejiang Industrial Co., Ltd, Hangzhou 310000, China

[2]College of Control Science and Engineering, Zhejiang University, Hangzhou 310027, China

[3]Windey Energy Technology Group Co, Hangzhou 310012, China

[4]DBAPPSecurity, Shanghai, 201306, China

[5]College of Electrical Engineering, Zhejiang University, Hangzhou 310027, China

*Corresponding author: Email: qyang@zju.edu.cn

**Abstract:** Industrial sensor networks exhibit heterogeneous, federated, large-scale, and intelligent characteristics due to the increasing number of Internet of Things (IoT) devices and different types of sensors. Efficient and accurate anomaly detection of sensor data is essential for guaranteeing the system's operational reliability and security. However, existing research on sensor data anomaly detection for industrial sensor networks still has several inherent limitations. First, most detection models usually consider centralized detection. Thus, all sensor data have to be uploaded to the control center for analysis, leading to a heavy traffic load. However, industrial sensor networks have high requirements for reliable and real-time communication. The heavy traffic load may cause communication delays or packets lost by corruption. Second, there are complex spatial and temporal features in industrial sensor data. The full extraction of such features plays a key role in improving detection performance. Nevertheless, the majority of existing methodologies face challenges in simultaneously and comprehensively analyzing both features. To solve the limitations above, this paper develops a cloud-edge collaborative data anomaly detection approach for industrial sensor networks that mainly consists of a sensor data detection model deployed at individual edges and a sensor data analysis model deployed in the cloud. The former is implemented using Gaussian and Bayesian algorithms, which effectively filter the substantial volume of sensor data generated during the normal operation of the industrial sensor network, thereby reducing traffic load. It only uploads all the sensor data to the sensor data analysis model for further analysis when the network is in an anomalous state. The latter based on GCRL is developed by inserting Long Short-Term Memory network (LSTM) into Graph Convolutional Network (GCN), which can effectively extract the spatial and temporal features of the sensor data for anomaly detection. The proposed approach is extensively assessed through experiments using two public industrial sensor network datasets compared with the baseline anomaly detection models. The numerical results demonstrate that the proposed approach outperforms the existing state-of-the-art models.

**Keywords:** Industrial sensor data, Gaussian and Bayesian algorithms, Anomaly detection, Long Short-Term Memory network, Graph convolutional network, Cloud-edge computing environment.

## Highlights

▶A framework for collaborative data anomaly detection is proposed for industrial sensor networks, which effectively reduces the traffic load.

▶A novel machine learning model is proposed, which can effectively extract the spatial and temporal features of industrial sensor data.

▶The proposed approach outperforms the existing models with an overall improvement of 4.35% for the F1-score.

## 1. Introduction

Industrial sensor networks have been developed quickly in the paradigm of the IoT and adopted in many industrial application domains ([1-3]) In industrial sensor networks, a large amount of sensor data, i.e., the measured value of sensors, is generally collected by different types of distributed industrial sensors and made available to a remote control center for centralized processing and analysis. The sensor data is valuable for condition monitoring and management functionalities in industrial sensor networks. However, anomalous sensor data cannot be completely avoided in practice due to many factors, e.g., unreliable sensing devices [4], computation, physical attacks, and transmission errors [5], imposing obvious risks to industrial sensor network applications. The failure to detect the anomalous speed of centrifuges led to the collapse of the nuclear plant, as documented in reference [6]. In the Venezuelan power grid event of 2019, anomalous voltage measurements were not accurately and timely detected, which resulted in a nationwide blackout, as described in reference [7]. Therefore, accurate anomalous sensor data detection is necessary to ensure the system's operational reliability ([8-15]).

At present, there are many industrial sensor data anomaly detection approaches, e.g., statistical algorithms and machine learning algorithms. However, most of them still exhibit two limitations, the first being issues with the detection architecture, and the second being challenges related to the extraction of features from industrial sensor data. Specifically, these limitations are as follows:

(1) Traditional sensor data anomaly detection models are based on statistical algorithms (e.g., [16-18]), which cannot provide high detection accuracy. Thus, machine learning algorithms are popular in sensor data anomaly detection due to their better detection performance (e.g., [19-22]). However, it should be noted that the machine learning-based model can be computationally complex for field sensors with limited computational capability. Therefore, many machine learning-based detection models are supposed to be deployed in the control center in practice. In Fig. 1, the arrows indicate the transmission of sensor data. As Fig. 1 (a) shows, the anomaly detection model using machine learning algorithms is deployed in the control center, so all sensor data have to be uploaded to the control center for further analysis. With the advancement of informatization and intelligence in Industrial Control Systems (ICS), the number of industrial sensors has drastically increased, and their distribution has become more widespread. Therefore, the centralized detection mechanism imposes a significant communication burden, leading to a surge in traffic load. However, reliable and real-time communication is essential to the stable operation of the industrial sensor network [23]. The increase in traffic load may disrupt normal communication behaviors ([24, 25]), leading to serious problems, e.g., communication delays and packets lost by corruption. Consequently, the traditional centralized detection architecture struggles to adapt to the evolving landscape of industrial control systems. Therefore, it is imperative to propose a new detection architecture to address this challenge.

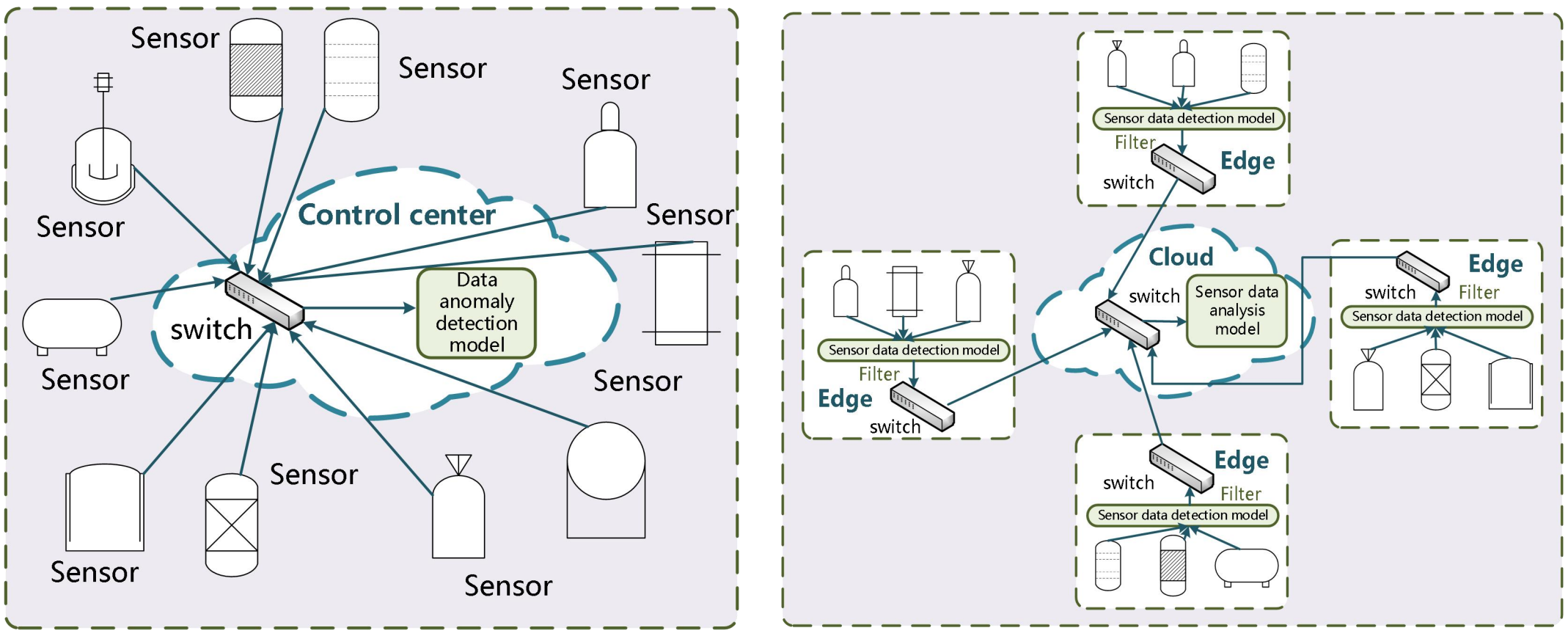

(a) Centralized data anomaly detection approach (b) Cloud-edge collaborative anomaly detection approach

Fig. 1. Different sensor data anomaly detection architecture

(2) Different industrial processes are generally coupled. Hence, the sensor data exhibit spatial and temporal features [26, 27]. In industrial sensor networks, the measurements of sensors at different positions influence each other. For example, a rise in water level may cause a drop in PH in a water treatment system. This is called a spatial feature. In addition, sensor data is time sequence data. Its value at time T+1 is affected by the value at T. This is called a temporal feature. It is crucial to consider spatial and temporal features simultaneously for sensor data characterization and anomaly detection [28]. In our investigation of existing literature on anomaly detection in sensor data within industrial sensor networks, we observed that the majority of studies focus on handling either temporal [29] or spatial [30] features. While there are a few articles capable of extracting both types of features simultaneously, their feature extraction capabilities remain insufficiently comprehensive.

To address limitation (1), we draw inspiration from the federated learning architecture ([31-33]), as well as our previous work on cloud–edge collaborative detection for network traffic anomaly detection [34], and further propose a cloud–edge collaborative detection framework tailored for sensor data anomaly detection. This approach takes into consideration that industrial sensor networks spend significantly more time in normal operation than in an anomalous state in real-world scenarios. This approach entails deploying the sensor data detection model at the edge and the sensor data analysis model in the cloud, as illustrated in Figure 1 (b). The sensor data detection model deployed at the edge efficiently filters the substantial volume of sensor data generated during normal operation of the industrial sensor network. It only uploads all the sensor data when the network is in an anomalous state to the cloud-based sensor data analysis model for further analysis. Therefore, this approach can lead to a significant reduction in traffic load. In this paper, we assume computational complexity as either high or low. High complexity is attributed to machine learning algorithms that require GPU involvement for computation, whereas low complexity is associated with statistical algorithms that do not require GPU support. Based on prior knowledge of ICS, it is understood that most edge areas in industrial sensor networks are low-computing environments, lacking GPU support, and therefore unable to perform deep learning operations. Therefore, the sensor data detection model is expected to be of low complexity. The Gaussian and Bayesian algorithms [35] based on statistical methods have low complexity and are widely used in anomaly detection applications. Thus, we select Gaussian and Bayesian algorithms as the sensor data anomaly detection model.

To address limitation (2), we introduce a sensor data analysis model called GCRL. This model combines the functionalities of GCN, Residual networks, and LSTM to analyze spatial and temporal features simultaneously.

GCN is popularized and widely used for extracting spatial features [36]. LSTM, with its well-established track record of effectively capturing temporal dependencies and patterns in sequential sensor data, has been widely adopted and extensively studied in the field of anomaly detection [37]. Therefore, we consider combining the two algorithms to mine temporal and spatial features of sensor data. The input of GCRL consists of a sensor data matrix and an adjacency matrix indicating the correlations between any two sensors. To construct the adjacency matrix, we use Spearman correlation analysis [38] to calculate the degrees of correlation between sensors. Moreover, to avoid the gradient disappearance or explosion during the training process, a Residual algorithm [39] is adopted in GCRL. The Residual algorithm is a technique aimed at mitigating the issue of gradient vanishing or steep gradient descent during the training of deep learning models. In essence, it involves conceptualizing a set of consecutive neural network layers as a module, where the input to the next module is equal to the sum of the current module's output and its original input. Finally, we can use the sensor data analysis model to further analyze the sensor data uploaded by the detection model at the edge and then accurately identify anomalies.

To evaluate the proposed detection approach, two public industrial sensor network datasets are used for evaluation. One is Water Distribution (WADI) dataset [19] and another is the Condition Monitoring of Hydraulic Systems (CMHS) dataset [40]. The two datasets are popular and widely used in industrial sensor data anomaly detection evaluation. The numerical results show that our approach performs better than other state-of-the-art anomaly detection approaches in detection performance. Compared with other anomaly detection approaches, our approach achieves an overall improvement of 4.35% for the *F1-score*. The main technical contributions made in this work are summarized as follows:

(1) A cloud-edge collaborative anomaly detection approach for industrial sensor networks is proposed, filtering the substantial volume of sensor data generated during normal operation of the industrial sensor network.

(2) A novel machine learning classification model GCRL is proposed to effectively mine the spatial and temporal features of industrial sensor data, which can improve detection accuracy.

(3) The proposed approach is extensively assessed through experiments based on two benchmark industrial sensor datasets. The results demonstrate that the proposed approach can achieve an impressive 4.35% improvement in *F1-score*, compared with the existing models.

The rest of the paper is organized as follows: Section II discusses the existing anomaly detection solutions in industrial sensor networks. The proposed cloud-edge collaborative anomaly detection approach for industrial sensor networks is presented in Section III. The experimental results of the proposed approach are compared with existing baseline approaches in Section IV. The conclusions and future research directions are given in Section V.

## 2. Related Work

In the domain of statistical-based industrial sensor network anomaly detection, extensive research has been undertaken. The author in [17] proposed a mixture of probabilistic PCA models for fault detection, which can separate the input space into some local regions and deploy the linear sensor anomaly diagnosis model in each region. A novel sparse PCA model can complete the task of localizing anomalies by analyzing a sparse low-dimensional space of anomalous data [41]. The author in [42] proposed a distributed anomaly detection technique based on the Seasonal Autoregressive Integrated Moving Average (SARIMA), considering the limited computing resources. However, it excels primarily in analyzing data exhibiting pronounced seasonality and periodicity. In [18], a data-driven detection approach was developed based on hidden Markov models for industrial sensor networks. Industrial sensor data contains complex features and correlations. However, statistical algorithms cannot effectively analyze these features and correlations due to their simple computing structure, leading to poor detection performance.

To improve detection accuracy, anomaly detection models based on machine learning algorithms are popular. In [43], the author introduced a framework for anomaly identification comprising a Convolutional Neural Network (CNN) and a two-stage LSTM-based Autoencoder. In [44], the author proposed a machine-learning approach to model normal sensor data to detect anomalous behaviors. In [45], the author proposed a sensor data anomaly classification algorithm leveraging Convolutional Neural Networks (CNN) with a Partially Observable Markov Decision Process (POMDP) model. The work in [29] proposed an LSTM-Gauss-NBayes approach that combined the LSTM with the Gaussian Bayes model for anomaly detection in the sensor network. However, this approach has a high false-positive rate in the presence of non-Gaussian distribution data. The author proposed a low-complexity model to detect anomalous sensor data in an industrial sensor network [46]. In [47], the author proposed an Extreme Learning Machine and Mutual Information (ELM-MI) based anomaly detection model. This model utilizes sensor data to compute temporal correlations by using the AutoCorrelation Function (ACF). Additionally, the Genetic Algorithm (GA) is used to solve multi-objective optimization problems. In [48], the authors developed a deep autoencoding Gaussian model to identify anomalous data. This model consists of a Gaussian mixture model and deep autoencoders. It should be noted that these anomaly detection models using machine learning algorithms can be computationally complex for the field sensors with limited computational capability and hence can be hardly deployed in the field sensors in practice. Therefore, most of the existing anomaly detection models using machine learning algorithms adopt centralized detection deployed in the control center, leading to a heavy traffic load in industrial sensor networks.

The industrial sensor data contains complex temporal and spatial features. Therefore, efficiently extracting these features is key to improving the accuracy and efficiency of anomaly detection. In the realm of detecting temporal features, many scholars have conducted extensive research. Many researchers have proposed methods to analyze the features. In [19], the normal sensor data was adopted to train a Generative Adversarial Network (GAN), and the discriminator based on the LSTM-Recurrent Neural Network (RNN) was used to extract temporal features and compute anomaly scores. However, the detection accuracy of this model is low in a benchmark dataset. The author used a GAN based on RNN to identify anomalous data [49]. The author [50] developed an anomaly detection framework called Multi-Time Scale Deep Convolutional Generative Adversarial Network (MTS-DCGAN). The goal of this framework is to efficiently extract temporal features of time-series data in Industrial Control Systems (ICS). In [51], the author developed an anomaly detection framework based on federated learning, combining CNN with LSTM. The model in [20] based on multi-head CNN-LSTM was used for detecting anomalies in sensor data. The anomaly detection algorithm in [52] also consists of a multi-head mechanism, CNN, and LSTM. In [53], the authors proposed a simple yet efficient method, i.e. Multiresolution Self-supervised Discriminative Network (MS$^2$D-Net), for anomalous sensor data detection. Many scholars have proposed cloud-edge-based anomaly detection methods [54]. The model [55] employs a cloud-edge computing architecture and uses a lightweight LSTM to extract features from sensor data for anomaly detection tasks. However, the LSTM does not adequately capture the spatial features of sensor data, particularly neglecting spatial characteristics. The authors [56] reduced the dimensionality of the sensor data and identified the anomalous data based on autoencoders. However, these models only focus on the temporal features without considering the spatial features of industrial sensor data.

In the extraction of spatial features in the industrial sensor network space, some scholars have conducted relevant research based on Graph Neural Network (GNN). The Graph Deviation Network (GDN) [30] was used to predict future data by graph attention-based forecasting and the absolute error was computed for evaluating the graph deviation score. In [57], a framework for sensor data anomaly detection was proposed, consisting of automatically learning a graph structure, graph convolutional, and transformer. However, these two models cannot efficiently extract temporal features as their calculations are conducted in the spatial domain, resulting in poor detection performance.

Some scholars have conducted relevant research on how to simultaneously extract spatial and temporal features. In [58], this paper presented a methodology for anomaly detection and classification of spatially correlated sensor data. The proposed approach integrated three techniques, namely Detrended Cross-Correlation Analysis (DCCA), Autoregressive Integrated Moving Average (ARIMA), and Forgetting Factor Iterative Data Capture Anomaly Detection (FFIDCAD). However, the utilized DCCA only considers the pairwise correlations between sensors and fails to fully capture the spatial features of the entire sensor network. Additionally, the implementation of ARIMA, being a statistical algorithm, may not effectively extract temporal features from long-term sequences. In [59], the author combined 1D-CNN and GCN to analyze the temporal and spatial features. The 1D-CNN utilizes convolutional operations primarily for extracting local features, where each convolutional kernel is crafted to focus on a specific subset of the data. However, this attribute might hinder its effectiveness in capturing long-range dependencies and global structural nuances. Therefore, existing studies have not fully succeeded in simultaneously extracting the spatial and temporal features of industrial sensor data.

## 3. Framework of Proposed Cloud-Edge Collaborative Data Anomaly Detection Approach

This paper proposes a cloud-edge collaborative data anomaly detection approach for industrial sensor networks. For this cloud-edge collaborative detection framework, we initially addressed three issues:

i) How does the edge determine the abnormal state of the industrial sensor network?

Considering the limited computational resources at the edges, the sensor data detection model is expected to be of low complexity. The Gaussian and Bayesian algorithms [35] based on statistical methods have low complexity and are widely used in anomaly detection applications. Thus, we select Gaussian and Bayesian algorithms as the sensor data anomaly detection model. In general, the distribution of anomalous sensor data differs from that of normal sensor data. Therefore, one approach is to separately compute the distributions of normal sensor data and anomalous sensor data. Then, by comparing how closely the sensor data to be detected aligns with these two distributions, we can determine whether the current sensor data is normal or anomalous. However, the sensor data detection model deployed at the edge can only identify if a particular sensor's value is abnormal. Our objective is to detect whether the entire industrial sensor network is in an abnormal state. It is unreasonable to identify the abnormal state of industrial sensor networks only by detecting the value of one sensor, which will lead to a high False Positive Rate (FPR). In practice, most sensors will be anomalous when the industrial sensor network is in an abnormal state. Therefore, we consider that the industrial sensor network is anomalous when the number of abnormal sensors exceeds the pre-defined sensitivity coefficient $e$, i.e., a threshold for the tolerable maximum number of anomalous sensors.

ii) If the edge can identify abnormal data, why is it necessary to upload the sensor data to the cloud for further analysis when the industrial sensor network is in an abnormal state?

Industrial sensor networks typically have the following data structure: with a sampling period of $a$, at each sampling time point, values from each sensor can be collected at the current moment. It's worth noting that currently available industrial sensor network datasets usually only label the operational state of the network as normal or anomalous. For example, if a physical attack occurs in the industrial sensor network from 9:10 to 9:20, then every sampling time point during these 10 minutes is marked as an anomalous state. The Gaussian and Bayesian algorithms require the prior calculation of probability density functions for normal and anomalous data for each sensor to perform the classification task. To compute the probability density function for anomalous data of a particular sensor, we need to utilize that sensor's anomalous data. However, in these publicly available industrial sensor network datasets, we only know whether the industrial sensor network is in an anomalous state at a given time but do not have information about specific sensors that are anomalous. Therefore, to calculate the probability density functions for anomalous data of each sensor, we can only consider every sensor's data in the training set as anomalous when the industrial sensor network is in an anomalous state. In practice, when an industrial sensor network experiences an anomaly, the majority of sensor data will exhibit anomalies, while a minority of sensor data remains unaffected and remains normal. Therefore, if we consider every sensor's data in the training set as anomalous when the industrial sensor network is in an anomalous state, some of these presumed anomalous sensor data may actually be normal. Consequently, the probability density functions calculated for the sensor's anomalous data may not be as accurate, potentially causing the sensor data detection model to misclassify some normal sensor data as anomalous. This, in turn, results in misclassifying an industrial sensor network in a normal state as anomalous, leading to a high FPR. Additionally, due to the low computational complexity of the sensor data detection model, the sensor data detection model can only perform binary classification and cannot identify the specific type of anomaly. Therefore, when the industrial sensor network is marked as anomalous, it is necessary to upload all sensor data from the current time to the cloud-based sensor data analysis model for further analysis. This process helps improve detection accuracy, and identify the specific type of anomaly.

iii) Why is it essential to upload all sensor data from the industrial sensor network during an abnormal state rather than just the data from the sensors flagged as abnormal?

In general, when an industrial sensor network is in an anomalous state, the majority of sensors within the network tend to be in an anomalous state, while a minority of sensors remain in a normal state. If we upload only the data from the sensors marked as anomalous by the sensor data detection model when the industrial sensor network is in an anomalous state, the structure of the sensor data uploaded to the cloud would not be consistent. For instance, there may be 60 anomalous sensors at one moment and 80 at another. Therefore, the sensor data received by the cloud may consist of 60 data points at one time and 80 data points at another time, resulting in inconsistent data structures. However, a non-uniform data structure hinders our ability to extract complex features, posing significant challenges for subsequent analysis. To address this issue, we have to upload the values of all sensors at the current moment. This ensures a consistent data structure at the cloud at all times, enabling us to comprehensively extract features for further analysis.

The proposed cloud-edge collaborative data anomaly detection approach consists of three models: a data preprocess model, a sensor data detection model, and a sensor data analysis model in the cloud. These models are illustrated in Fig. 2, where the blue arrows represent the transmission of sensor data between different physical devices and various detection models; the red arrows denote trigger signals; the purple arrows represent the transmission of the number of anomalous sensors from the edge areas to the cloud; and the green arrows indicate the flow of sensor data processing within the detection models. The data preprocess model and sensor data detection model are deployed at individual edges, covering a collection of sensors. The data preprocess model can aggregate sensor data and transform non-Gaussian distributed data into Gaussian distributed data. The sensor data detection model can identify the sensors with anomalous data. The count of anomalous sensors in each edge is subsequently transmitted to the cloud, where the total number of anomalous sensors can be calculated by aggregating the information received from all edges. If the total number of anomalous sensors exceeds the predefined sensitivity coefficient $e$ (i.e., a threshold for the tolerable maximum number of anomalous sensors), the industrial sensor network is considered to be in an anomalous operational state. At this point, the data uploading mechanism in each edge will be triggered, prompting the transmission of sensor data from all edges to the cloud for further analysis. Thus, massive sensor data during normal operation of the industrial sensor network can be filtered, which makes the traffic load smaller. In the cloud, the sensor data analysis model based on GCRL comprises Spearman correlation analysis, GCN, LSTM, and a Residual algorithm. The core goal of GCRL is to efficiently extract and analyze the spatial and temporal features of industrial sensor data, ensuring accurate anomaly detection.

Without the data preprocessing model, the sensor data detection model would not function effectively. Sole reliance on the sensor data detection model for anomaly detection tasks could lead to low detection accuracy and challenges in identifying anomaly types. Furthermore, if the anomaly detection tasks are solely conducted by the sensor data analysis model deployed in the cloud, it would require continuous uploading of all sensor data generated by the industrial sensor network. Although this approach effectively extracts temporal-spatial features of the sensor data and yields higher detection accuracy, it significantly increases the traffic load, disrupts the normal communication behaviors of the ICS, and impedes their stable operation. Consequently, based on this analysis, the approach proposed in this paper is an integrated approach composed of three interdependent models. These models cannot function independently for anomaly detection; their roles are complementary and essential to one another.

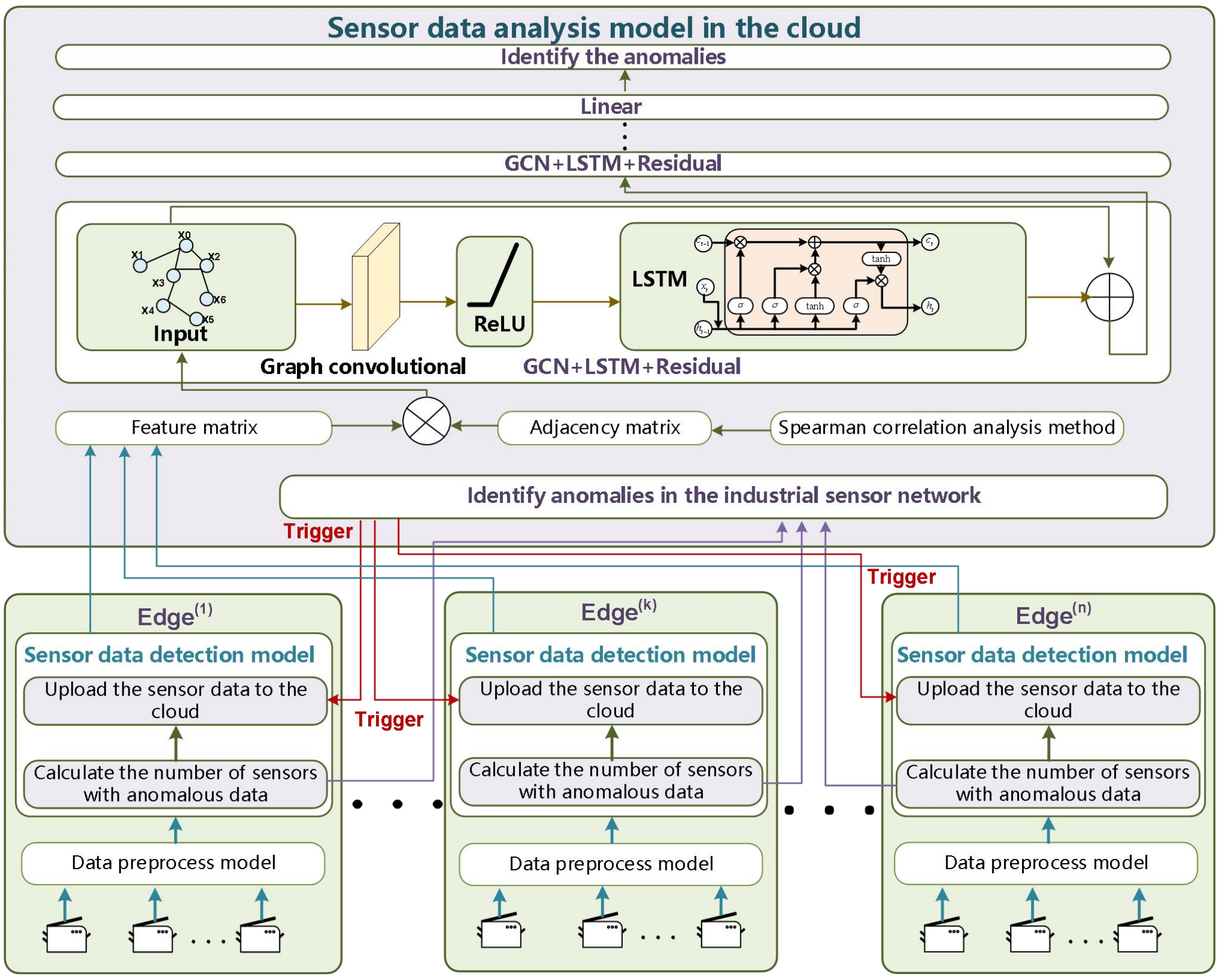

Fig. 2. Framework of the proposed cloud-edge collaborative data anomaly detection approach

In summary, the cloud-edge collaborative data anomaly detection approach we propose addresses two challenges present in current industrial sensor network anomaly detection. The first challenge is that existing detection frameworks primarily employ centralized detection, which can lead to increased traffic load. The second challenge is that current methods struggle to simultaneously extract the temporal and spatial features of industrial sensor networks comprehensively. The details of our proposed approach are presented in the following sections.

### *3.1 Data Preprocess Model*

The data preprocess model is designed to aggregate sensor data and convert non-Gaussian distributed data into Gaussian distributed data. In practice, the data sampling rate is generally high, e.g., in seconds or milliseconds. Thus, the data aggregation algorithm needs to be adopted for data analysis in multiple time scales. Moreover, The aggregated sensor data may not follow the Gaussian distribution, so the Box-Cox transformation [48] is used to transform the non-Gaussian distribution data into a Gaussian distribution. The details of the data preprocess model are given in equations (1)-(2).

$$X_{r,i,t} = x_{r,i,t} + x_{r,i,t-a} + \cdots + x_{r,i,t-Ba} \tag{1}$$

$$X'_{r,i,t} = \begin{cases} \dfrac{X_{r,i,t} - 1}{\lambda_{r,i}} \ \text{if}\ \lambda_{r,i} \neq 0 \\ \log[X_{r,i,t}]\ \text{if}\ \lambda_{r,i} = 0 \end{cases} \tag{2}$$

where $x_{r,i,t}$ represents the collected original data of sensor *i* in the *r*-th edge at time *t*; *a* is the sampling period and *B* is the scale of aggregation; $X_{r,i,t}$ is the aggregated data of sensor *i* in the *r*-th edge at time *t*. $\lambda_{r,i}$ is used to maximize the log-likelihood function of sensor *i* in the *r*-th edge. $X'_{r,i,t}$ is the preprocessed data of sensor *i* in the *r*-th edge at time *t*.

### *3.2 Sensor Data Detection Model at the Edge*

To mitigate traffic load, we have embraced a cloud-edge collaborative detection architecture that employs a sensor data detection model deployed at the edge to pre-filter substantial amounts of sensor data during the regular operation of the industrial sensor network. However, given the limited computing resources at the edge, the computational structure of the sensor data detection model cannot be overly complex. Consequently, we opted for a straightforward statistical algorithm for this model, namely the Gaussian and Bayesian algorithms. The detailed computational process of the sensor data detection model is outlined as follows:

First, we need to calculate the Probability Density Function (PDF) of normal and abnormal data. The dataset is split into two sets: a training set and a testing set, in which the label value 0 represents abnormal and 1 represents normal. We use the training set to calculate the PDF of normal and abnormal data. The details are shown in equations (3)-(4).

$$\begin{cases} f_{r,i,no} = \dfrac{1}{\sqrt{2\pi}\delta_{r,i,no}} \exp\left( -\dfrac{(X'_{r,i,t} - \mu_{r,i,no})^2}{2\delta^2_{r,i,no}} \right) \\ f_{r,i,ab} = \dfrac{1}{\sqrt{2\pi}\delta_{r,i,ab}} \exp\left( -\dfrac{(X'_{r,i,t} - \mu_{r,i,ab})^2}{2\delta^2_{r,i,ab}} \right) \end{cases} \tag{3}$$

where $f_{r,i,no}$ and $f_{r,i,ab}$ represents the normal and abnormal PDF of sensor $i$ in the $r$-th edge, respectively; $\mu_{r,i,no}$ and $\delta_{r,i,no}$ are the means and variance of the $i$-th sensor in the $r$-th edge when its label is 1. Otherwise, it is $\mu_{r,i,ab}$ and $\delta_{r,i,ab}$, respectively.

The parameters of normal and abnormal PDFs can be obtained through the maximum likelihood estimate. The parameters of PDF are shown in equation (4):

$$\begin{cases} \mu_{r,i,no} = \overline{X'_{r,i,no}} \\ \mu_{r,i,ab} = \overline{X'_{r,i,ab}} \\ \delta_{r,i,no} = \frac{1}{n_1}\sum_{j=1}^{n_1}(X'_{r,i,j,no} - \overline{X'_{r,i,no}})^2 \\ \delta_{r,i,ab} = \frac{1}{n_0}\sum_{j=1}^{n_0}(X'_{r,i,j,ab} - \overline{X'_{r,i,ab}})^2 \end{cases} \tag{4}$$

where $\overline{X'_{r,i,no}}$ and $\overline{X'_{r,i,ab}}$ represents the means in normal and abnormal training sets of $i$-th sensor data in the $r$-th edge, respectively. $X'_{r,i,j,ab}$ and $X'_{r,i,j,no}$ represents the preprocessed $j$-th abnormal and normal data of sensor $i$ in the $r$-th edge, respectively. $n_0$ and $n_1$ are the number of preprocessed abnormal and normal training data of sensor $i$ in the $r$-th edge, respectively.

Then, the normal and abnormal probability of the sensor data can be calculated based on equation (5).

$$\begin{cases} p(y=1 \mid X'_{r,i,j}) = \dfrac{p(X'_{r,i,t} \mid y=1)p(y=1)}{p(X'_{r,i,t} \mid y=1)p(y=1) + p(X'_{r,i,t} \mid y=0)p(y=0)} \\ p(y=0 \mid X'_{r,i,j}) = \dfrac{p(X'_{r,i,t} \mid y=0)p(y=0)}{p(X'_{r,i,t} \mid y=1)p(y=1) + p(X'_{r,i,t} \mid y=0)p(y=0)} \end{cases} \tag{5}$$

As $p(X'_{r,i,j} \mid y=1)p(y=1) + p(X'_{r,i,j} \mid y=0)p(y=0)$, i.e., the denominator is the same when $p(y=1 \mid X'_{r,i,j})$ and $p(y=0 \mid X'_{r,i,j})$ are compared. Thus, equation (5) can be further simplified as equation (6).

$$\begin{cases} p(y=1 \mid X'_{r,i,t}) = f_{r,i,no}(X'_{r,i,t})p(y=1) \\ p(y=0 \mid X'_{r,i,t}) = f_{r,i,ab}(X'_{r,i,t})p(y=0) \end{cases} \tag{6}$$

where $p(y=1)$ and $p(y=0)$ are constants, which represent the proportion of normal and abnormal samples in the training set, respectively.

The algorithm of the sensor data detection model is given in Algorithm 1.

| Algorithm 1: Sensor data detection model at the edge |
| --- |
| Input: the training set, real-time original sensor data $x_{r,i,t}$, $r = (1,2,\cdots,k)$, $i = (1,2,\cdots,m_r)$<br>Output: all sensor data $F_t$ at the current moment |
| 1: **Aggregate** the original sensor data using the training set<br>2: **Transform** the non-Gaussian distribution into Gaussian distribution using the equation (2)<br>3: **Obtain** the normal and abnormal PDFs for each sensor using the equation (4)<br>4: **Collect** sensor data<br>5: $s$=0<br>6: $X_{r,i,t} = x_{r,i,t} + x_{r,i,t\text{-}a} + \cdots + x_{r,i,t\text{-}Ba}$ |

| | |
|---|---|
| 7: | Transform the data into Gaussian distribution |
| 8: | Compute the $p(y=1 \mid X'_{r,i,t})$ . and $p(y=0 \mid X'_{r,i,t})$ |
| 9: | Compare the $p(y=1 \mid X'_{r,i,t})$ and $p(y=0 \mid X'_{r,i,t})$ |
| 10: | **If** $p(y=0 \mid X'_{r,i,t}) \geq p(y=1 \mid X'_{r,i,t})$ |
| 11: | \| $s$=$s$+1 |
| 12: | **If** $s > e$ |
| 13: | \| upload all the sensor data $F_t$ at the current moment to the cloud |
| 14: | **Jump** *step 4* |

where $k$ is the number of edge areas in the industrial sensor network; $m_r$ is the number of sensors in the $r$-th edge; $s$ is the number of anomalous sensors in the whole sensor network.

Not all sensor data can be transformed into Gaussian distribution using the Box-Cox algorithm. Therefore, the $p(y=1 \mid X'_{r,i,t})$ and $p(y=0 \mid X'_{r,i,t})$ may be equal when the transformed data do not follow Gaussian distribution. The sensor data can't be identified as normal or abnormal in this case, so we consider the sensor data as potentially abnormal and in need of further analysis. Therefore, we set this condition $p(y=0 \mid X'_{r,i,t}) \geq p(y=1 \mid X'_{r,i,t})$ . It is unreasonable to identify the abnormal state of industrial sensor networks only by detecting the value of one sensor, which will lead to a high FPR. In practice, most sensors will be anomalous when the industrial sensor network is in an abnormal state. Therefore, we consider that the industrial sensor network is anomalous when the number of abnormal sensors exceeds the pre-defined sensitivity coefficient $e$.

### *3.3 GCRL-based Sensor Data Analysis Model in the Cloud*

Given the inherent correlation among data from various sensors in coupled industrial processes, the sensor data inherently manifests both spatial and temporal features. To comprehensively extract these dual features, we introduce GCRL, a framework comprising GCN, LSTM, and a Residual algorithm. GCN is used to extract the spatial features of sensor data. The input of GCN consists of a sensor data matrix and an adjacency matrix indicating the correlations between any two sensors. In fact, any two sensors have different degrees of correlation in the industrial sensor network. To construct the adjacency matrix, we use the Spearman correlation analysis algorithm to calculate different degrees of correlation between any two sensors. Moreover, to extract temporal features, we insert LSTM into GCN to process sensor data. In addition, a Residual algorithm is also adopted in the proposed GCRL to avoid gradient disappearance or explosion. The implementation of the proposed GCRL in the cloud is described as follows:

Firstly, the adjacency matrix, representing the correlations between any two sensors in the sensor network is built using the Spearman correlation analysis algorithm. The details of the calculation are shown in equations (7) and (8).

$$\rho_{z,c} = 1 - \frac{6\sum_{u=1}^{n_{z,c}} d^2_{z,c,u}}{{n_{z,c}}^3 - n_{z,c}} (z \neq c) \tag{7}$$

$$p_{z,c} = (1 - 2 * \int_{-\infty}^{\rho_{z,c}} \frac{1}{\sqrt{2\pi}} \exp(-x^2/2)) \tag{8}$$

where $\rho_{z,c}$ is the correlation coefficient between sensor $z$ and sensor $c$. $d_{z,c,u}$ is the difference between the $u$-th pair values in the dataset that represents the sorted data of the value of sensor $z$ and sensor $c$. $n_{z,c}$ is the number of the value of sensor $c$ or sensor $z$. $p_{z,c}$ is the P-value [60] for the $\rho_{z,c}$, which is used to determine whether there exists a correlation between the value of sensor $z$ and sensor $c$.

The adjacency matrix can be calculated using equation (9).

$$\begin{cases}(z,c)\in A & \rho_{z,c} > p \text{ and } p_{z,c} < \alpha \\ (z,c)\notin A & \rho_{z,c} < p \text{ or } p_{z,c} > \alpha\end{cases} \tag{9}$$

where $z$ or $c$ represents the sensor number. $p$ is the threshold of the correlation coefficient and $\alpha$ is the significance level. $A$ is the adjacency matrix of the industrial sensor network.

Each layer in the GCRL model consists of a graph convolutional part, LSTM, and residual part.

$$y_{G,l+1} = f_{G,l+1}(y_{L+G,l}, A) = \delta(\hat{D}^{-\frac{1}{2}}\hat{A}\hat{D}^{-\frac{1}{2}} y_{L+G,l} W^{(l+1)}) \tag{10}$$

$$y_{L,l+1} = \text{LSTM}_{l+1}(y_{G,l+1}) \tag{11}$$

$$y_{L+G,l+1} = y_{L,l+1} + y_{L+G,l} \tag{12}$$

where $\delta$ is the activation function; $\hat{A} = A + I$ represents the adjacency matrix with self-loops; $\hat{D}$ is the corresponding degree matrix. $W^{(l+1)}$ is the weight matrix of the graph convolutional part in $l$+1-th layer. $y_{G,l+1}$ is the output of the graph convolutional part in $l$+1-th layer. $\text{LSTM}_{l+1}()$ represents the LSTM function in the $l$+1-th layer. $y_{L,l+1}$ is the result of the LSTM part in the $l$+1-th layer. $y_{L+G,l+1}$ is the final output of $l$+1-th layer. Moreover, $y_{G,1} = f_{G,1}(F_t, A) = \delta(\hat{D}^{-\frac{1}{2}}\hat{A}\hat{D}^{-\frac{1}{2}} F_t W^{(1)})$ and $F_t$ is the input of the GCRL that is provided by the sensor data detection model at the edges.

The structure of the proposed sensor data analysis model is shown in Fig. 3, where the blue arrows represent the transmission of sensor data between different physical devices and various detection models; and the green arrows indicate the flow of sensor data processing within the detection model.

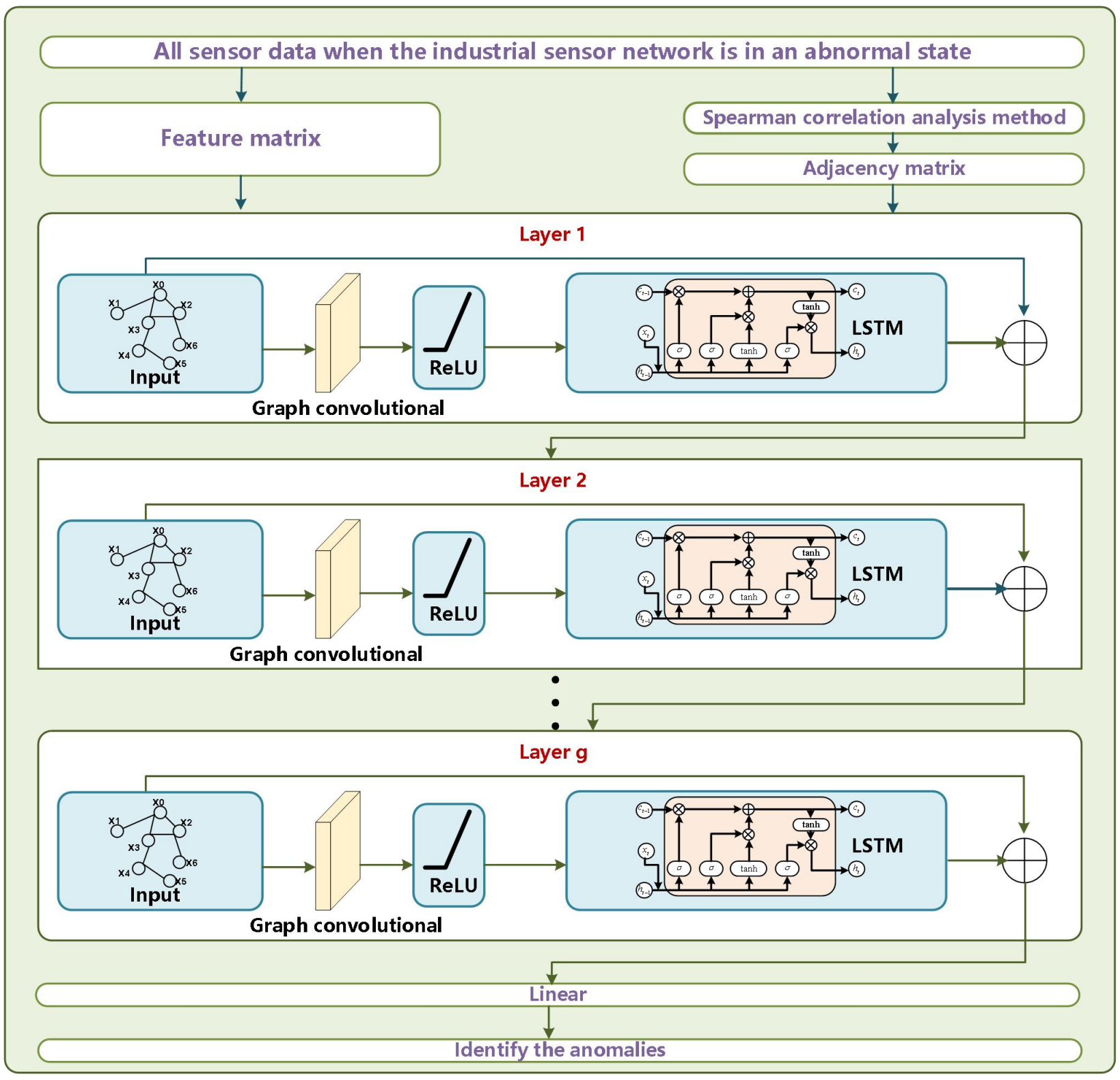

Fig. 3. Computing structure of the proposed sensor data analysis model

### 3.4 The Cloud-edge Collaborative Data Anomaly Detection Approach

The proposed cloud-edge collaborative data anomaly detection approach can identify the abnormal state of an industrial sensor network and upload the values of all sensors at the current moment. Consequently, this approach helps reduce the transmission of massive sensor data during the normal operation of the industrial sensor network to the cloud, thereby reducing the traffic load. The sensor data analysis model based on GCRL deployed in the cloud can sufficiently extract the spatial and temporal features, improving detection performance. The algorithm of the proposed approach is given in Algorithm 2.

---

Algorithm 2: Cloud-edge collaborative data anomaly detection approach

In: the training set, real-time original sensor data $x_{r,i,t}$ , $r = (1, 2, \cdots, k)$ , $i = (1, 2, \cdots, m_r)$

Out: classification result

---

1: **Obtain** the normal and abnormal PDFs for each sensor using the equations (1)-(4)
2: **Calculate** the adjacency matrix $A$ using the training set. using equation (7)-(9)
3: **Train** the GCRL model using the training set.
4: **Obtain** original sensor data $x_{r,i,t}$ at the edges
5: **If** the state of the sensor network is identified as normal using the sensor detection model at the edges
6: | **Jump** *Step 4*
7: **Else**
8: | Upload all sensor data $F_t$ at the current moment to the cloud
9: | $A$ and $F_t$ are the input of the GCRL model
10: | **for** $l \leftarrow 1$ to $g$ do
11: | | Extract the spatial features using equation (10)
12: | | Process the temporal features using equation (11)
13: | | Obtain the output of layer $l$ using the equation (12)
14: | **Obtain** the classification result
15: **Jump** *Step 4*

---

where $g$ is the number of layers of the sensor data analysis model based on GCRL.

## 4. Testbed Experiment and Numerical Results

The proposed cloud-edge collaborative data anomaly detection approach is assessed using the Water Distribution (WADI) dataset [19] and Condition Monitoring of Hydraulic Systems (CMHS) dataset [40], compared with a range of existing detection models. At each edge area, a host machine equipped with an Intel(R) Core(TM) i5-9400 CPU and 4GB of memory is deployed for computation, without a GPU graphics card. The CPU operates on Ubuntu 20.04.6 LTS. In the cloud, a deep learning server is configured, with details of its hardware environment provided in Table 1. The sensor data analysis model is trained using a significance level of $\alpha = 0.05$ and ReLu is adopted as the activation function. The specific structure and parameters of the deep learning algorithm can be explored in our publicly available source code.

Table 1. The details of the hardware

| | |
|---|---|
| Motherboard | MSI PRO X570-A PRO motherboard |
| CPU | Intel(R) Core (TM) i9-9820X CPU |
| Memory | Kingston FURY 64GB DDR4 3200 |
| Storage | WD40EZAZ 4TB SATA |
| GPU | NVIDIA RTX 3090 GPU |
| Power supply | Great Wall Dragon e-sports version 800W |

### 4.1 Dataset

#### 4.1.1 WADI dataset

The WADI testbed [19] is a water distribution system consisting of many pipelines and 127 sensors spanning across a large area, forming a storage and distribution network, and realistic and complete water treatment. The WADI dataset consists of 14 types of attack data, e.g., turn on or turn off maliciously some actuators, stop chemical dosing to the raw water, and 14-day continuous data in normal operational conditions. During the normal and attack data collection, all sensor data were collected. The sampling period is 1000ms, representing 127 sensor data that can be collected with the sampling rate of 1000ms. In this work, we define the values collected from all sensors per second as one sample. If a sample is labeled as anomalous, it indicates that at the corresponding time, the industrial sensor network is in an abnormal operational state. The details of 15-classification WADI dataset is shown in Table 2. To evaluate the binary classification performance of the proposed approach, we consider the 14 types of attack data as abnormal data and then obtain a 2-classification WADI dataset, as shown in Table 3. The values in these two tables represent the number of samples collected for each item. For example, 1188 in Table 2 indicates that the number of samples of attack #2 is 1188 for training.

Table 2

Details of 15-classification WADI dataset

| Item | Training | Testing |
|---|---|---|
| normal #0 | 129210 | 33614 |
| attack #1 | 1188 | 298 |
| attack #2 | 461 | 115 |
| attack #3 | 1383 | 345 |
| attack #4 | 670 | 168 |
| attack #5 | 522 | 130 |
| attack #6 | 548 | 138 |
| attack #7 | 455 | 113 |
| attack #8 | 70 | 18 |
| attack #9 | 636 | 158 |
| attack #10 | 527 | 132 |
| attack #11 | 279 | 70 |
| attack #12 | 153 | 38 |
| attack #13 | 455 | 113 |
| attack #14 | 496 | 124 |

Table 3

Details of 2-classification WADI dataset

| Item | Training | Testing |
|---|---|---|
| Normal | 129210 | 33614 |
| Abnormal | 7843 | 1960 |

### 4.1.2 CMHS dataset

The CMHS dataset [40] was obtained from an industrial hydraulic test rig that is composed of a primary working, a secondary cooling-filtration circuit, and an industrial sensor network. There are 17 sensors in the industrial sensor network (e.g., temperature, pressure, motor power), and each sensor obtains a measurement per second. The 17 sensor data collected per second is defined as one sample. The dataset has two labels: cooler condition and valve condition. The cooler condition has three states: 3 is "close to total failure", 20 is "reduced efficiency", and 100 is "full efficiency". The valve condition has four states: 73 is "close to total failure", 80 is "severe lag", 90 is "small lag", and 100 is "optimal switching behavior". The details of 6-classfication CMHS dataset is shown in Table 4. To evaluate the binary classification performance of the proposed approach, we consider the condition where a cooler condition value of 3 is treated as abnormal, and a value of 100 is treated as normal. Based on this criterion, we construct a binary-class WADI dataset, as summarized in Table 5. The values in the two tables represent the number of samples.

Table 4

Details of 6-classification CMHS dataset

| Item | Training | Testing |
|---|---|---|
| attack @0: cooler condition is 3 and valve condition is 73 or 80 | 11520 | 2880 |
| attack @1: cooler condition is 3 and valve condition is 90 or 100 | 23616 | 5900 |
| attack @2: cooler condition is 20 and valve condition is 73 or 80 | 11520 | 2880 |
| attack @3: cooler condition is 20 and valve condition is 90 or 100 | 23616 | 5904 |
| attack @4: cooler condition is 100 and valve condition is 73 or 80 | 11520 | 2880 |
| normal @5: cooler condition is 100 and valve condition is 90 or 100 | 24048 | 6012 |

Table 5

Details of 2-classification CMHS dataset

| Item | Training | Testing |
|---|---|---|
| Abnormal: cooler condition is 3 | 35136 | 8780 |
| Normal: cooler condition is 100 | 35568 | 8892 |

### 4.2 Detection Performance at the Edge

To evaluate the detection performance of the proposed approach at the edge, we use the 2-classification WADI dataset and 2-classification CMHS dataset to conduct this experiment. $B$ is selected as 10, representing that the sensor data is detected every 10 seconds. The evaluation metrics are shown as follows:

$$FR = \frac{FP}{TN + FP} \tag{13}$$

$$k_times = TN + FN \tag{14}$$

$$RTL=(N_{ab+no} - k_times) * B * N_sensors \tag{15}$$

where True Negative (*TN*) represents the number of abnormal samples classified as abnormal correctly. False Positive (*FP*) is the number of abnormal samples classified as normal incorrectly. False Positive (*FN*) represents the number of normal samples classified as abnormal incorrectly. *FR* represents the proportion of abnormal samples that are classified as normal samples. $k_times$ is the times of uploading the anomalous samples to the cloud. *RTL* represents the number of filtered sensor data by the sensor data detection model, which is corresponding to the traffic load. $N_{ab+no}$ is the number of samples in the testing set. $N_sensors$ is the number of sensors in the industrial sensor network.

This work also considers the influence of sensitivity coefficient *e* on the detection results. The numerical results are presented in Table 6 and Table 7.

Table 6

Results of the sensor data detection model using the 2-classification WADI dataset

| *e* | *TN* | *FN* | *FR* | *k_times* | *RTL* |
|---|---|---|---|---|---|
| 15 | 196 | 3361 | 0 | 3557 | 0 |
| 20 | 196 | 3037 | 0 | 3233 | 411480 |
| 21 | 196 | 2464 | 0 | 2660 | 1139190 |
| 22 | 196 | 1374 | 0 | 1570 | 2523490 |
| 23 | 196 | 1012 | 0 | 1208 | 2983230 |
| 24 | 196 | 843 | 0 | 1039 | 3197860 |
| 25 | 196 | 759 | 0 | 955 | 3304540 |
| 26 | 196 | 721 | 0 | 917 | 3352800 |
| 27 | 196 | 663 | 0 | 859 | 3426460 |
| 28 | 193 | 538 | 1.53% | 731 | 3589020 |
| 29 | 167 | 243 | 14.79% | 410 | 3996690 |
| 31 | 111 | 0 | 43.37% | 111 | 4376420 |
| 33 | 19 | 0 | 90.31% | 19 | 4493260 |

Table 7

Results of the sensor data detection model using the 2-classification CMHS dataset

| *e* | *TN* | *FN* | *FR* | *k_times* | *RTL* |
|---|---|---|---|---|---|
| 1 | 878 | 889 | 0 | 1767 | 0 |
| 2 | 878 | 889 | 0 | 1767 | 0 |
| 3 | 878 | 183 | 0 | 1061 | 120020 |
| 4 | 878 | 148 | 0 | 1026 | 125970 |
| 5 | 878 | 4 | 0 | 882 | 150450 |
| 6 | 854 | 0 | 2.73% | 854 | 155210 |
| 7 | 644 | 0 | 26.66% | 644 | 190910 |
| 8 | 226 | 0 | 74.26% | 226 | 261970 |
| 9 | 0 | 0 | 100% | 0 | 300390 |

Table 6 shows that as $e$ increases, $TN$ and $FN$, as well as $k_times$, will decrease, while $FR$ and $RTL$ will increase. On the one hand, we want to reduce the traffic load and minimize $k_times$ while maximizing $RTL$. On the other hand, the sensor data detection model cannot identify the specific type of anomaly. Thus, all anomalous samples must be uploaded to the cloud for further analysis, i.e. $FR$ must be zero. Therefore, it seems that $e$=27 is the best choice. However, the proposed approach consists of the sensor data detection model and the sensor data analysis model. Different values of $e$ can affect the classification results of the sensor data analysis model in the cloud. Thus, the detection performance of the sensor data analysis model may not be the best when $e$=27. We need to consider $e$ with different values for further analysis. It can be observed that $RTL$ when $e$=22 is about twice as much as $RTL$ when $e$=21, so we consider $e$=22 as a turning point. Moreover, $e$=27 is the performance turning point as $FR$ is not zero when $e$ is more than 27. Therefore, we conduct the following experiments using $e$ with the different values of 22, 23, 24, 25, 26, and 27.

In Table 7, we can find that all normal samples will be uploaded to the cloud for further analysis when $e$=2. When $e$=3, most of the normal samples will be filtered and the sensor data detection model can reduce 120020 sensor data delivered to the cloud. Thus, we consider $e$=2 is a turning point. In addition, We observed a sharp increase in $FR$ when transitioning from $e$ =7 to $e$ =8. Thus, we consider $e$ =7 is a turning point. In the following experiment, we consider $e$ with different values of 2, 3, 4, 5, 6, and 7 for further analysis.

The results in Table 6 and Table 7 demonstrate that if we select an appropriate value for $e$, the sensor data detection model can accurately identify all anomalous states of the industrial sensor network and upload the data from all sensors at the current anomalous time. Additionally, it can filter out a significant amount of sensor data during the normal operation of the industrial sensor network, reducing the traffic load.

### 4.3 Detection Performance in the Cloud

In this section, we conducted four experiments to evaluate the proposed anomaly detection approach in the following four aspects.

Experiment A: What is the influence of different parameter settings on detection results? How to select the parameters with the best detection performance?

Experiment B: Whether the proposed approach can learn spatial and temporal features of industrial sensor data, by inserting LSTM into GCN?

Experiment C: How is the performance of our proposed detection approach compared with other sensor data anomaly detection approaches for binary classification?

Experiment D: How is the performance of our proposed detection approach compared with other detection approaches for the classification of multiple anomalies?

#### 4.3.1 Experiment A: The Detection Results of Different Parameter Settings

Experiment A primarily addresses the following questions: What is the influence of different parameter settings on the detection results? How can the parameters that yield the best detection performance be selected?

In Experiment A, the datasets in Table 3 and Table 5 are used for evaluation. First, we construct the adjacency matrix $A$ using the training set and the Spearman correlation analysis algorithm. The threshold of correlation coefficient $p$ is 0.45, 0.55 or 0.65. The corresponding sensor data correlation graph is used to visualize adjacency matrix $A$, as shown in Fig. 4 and Fig. 5. Each point represents a sensor, and the coordinates are calculated using the T-distributed Stochastic Neighbor Embedding (TSNE) [61]. If one sensor is associated with other sensors, the sensor is a red point, while the blue points are the opposite. The edge represents that there exists a correlation between the two sensors. The edges in Fig. 4(a) are denser than those in Fig. 4(b) and Fig. 4(c), indicating that the sensor data of WADI is sensitive to the value of $p$. In Fig. 5, we find that the structures of the three correlation graphs are different but the density is similar, demonstrating that the sensor data of CMHS is not sensitive to the value of $p$. The adjacency matrix $A$ is the input of the proposed approach, and the threshold of correlation coefficient $p$ has an influence on adjacency matrix $A$. Therefore, the parameter $p$ will affect the final classification results.

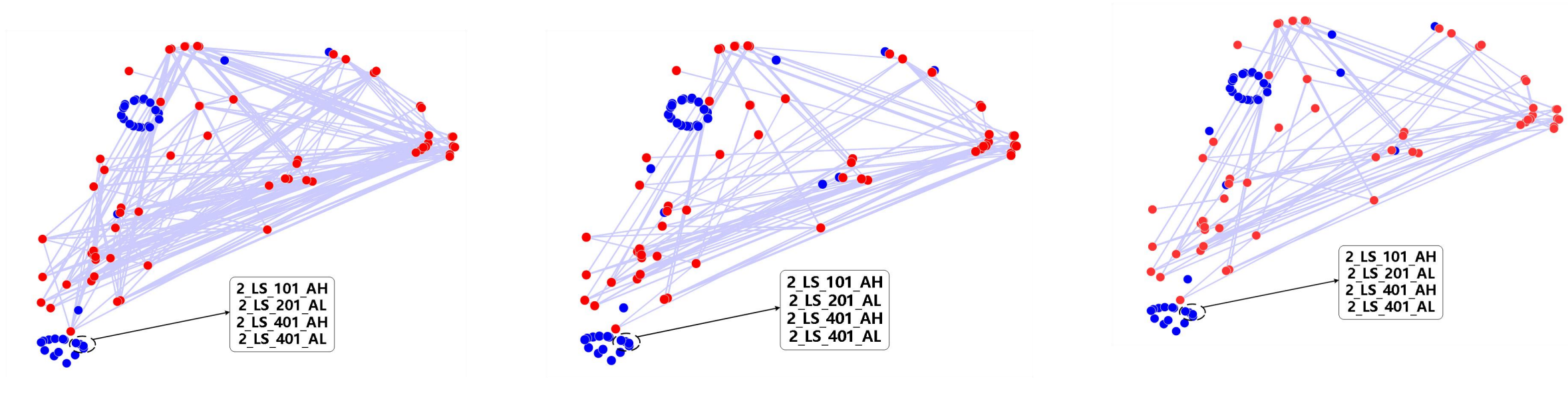

(a) $p$ = 0.45 (b) $p$ = 0.55 (c) $p$= 0.65

Fig. 4. Sensor data correlation graphs of WADI

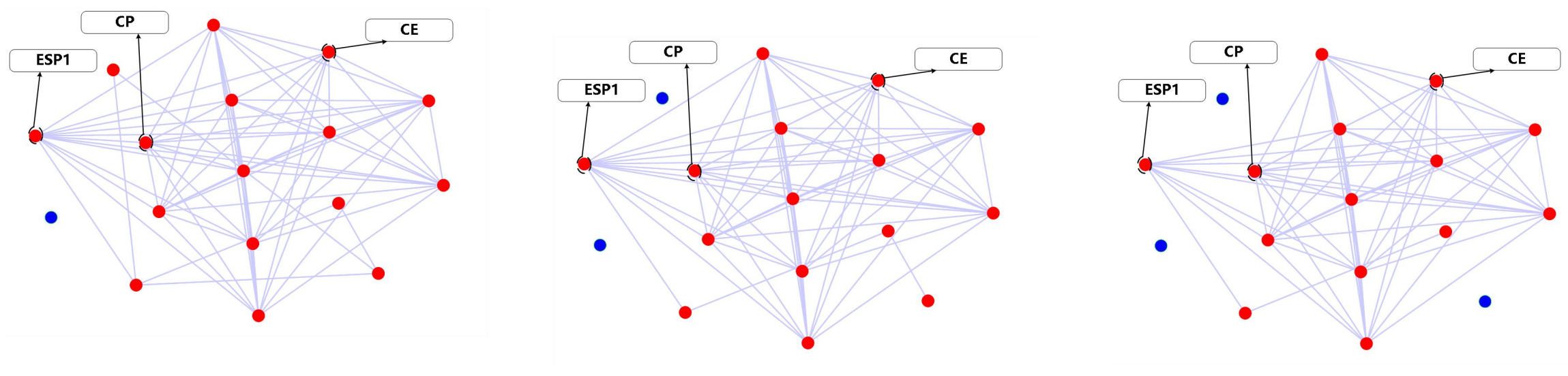

(a) $p$ = 0.45 (b) $p$ = 0.55 (c) $p$= 0.65

Fig. 5. Sensor data correlation graphs of CMHS

Then, the proposed approach is evaluated with different parameter settings. Here, the parameter $e$ is 22, 23, 24, 25, 26, 27 for the WADI dataset, and $e$ is 2, 3, 4, 5, 6, 7 for the CMHS dataset. The parameter $p$ is 0.45, 0.55, or 0.65.

Three commonly used classification metrics, namely *Precision*, *Recall,* and *F1-score*, are employed for performance assessment.

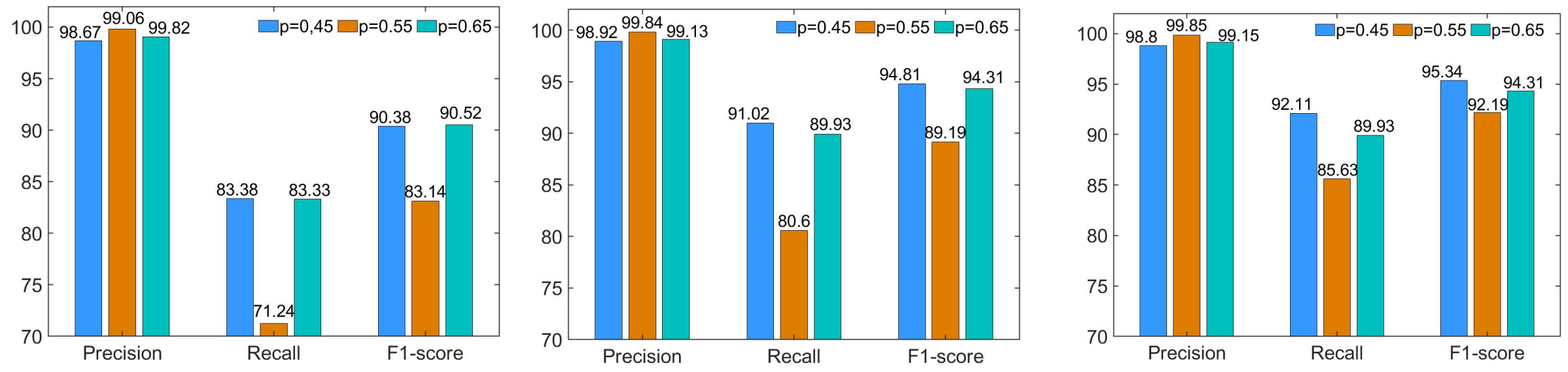

(a) $e$=22 (b) $e$=23 (c) $e$=24

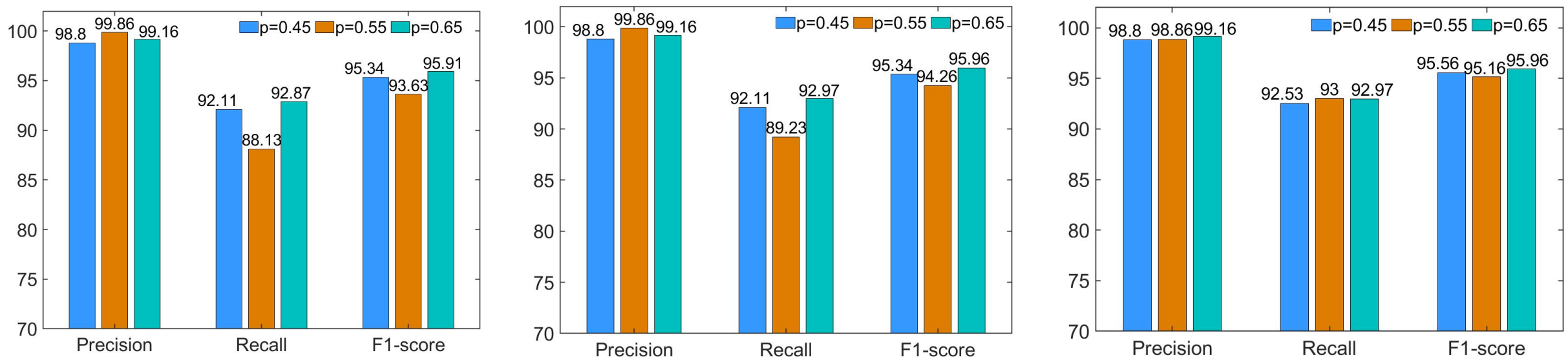

(d) $e$=25 (e) $e$=26 (f) $e$=27

Fig. 6. Comparison of different parameter settings (WADI)

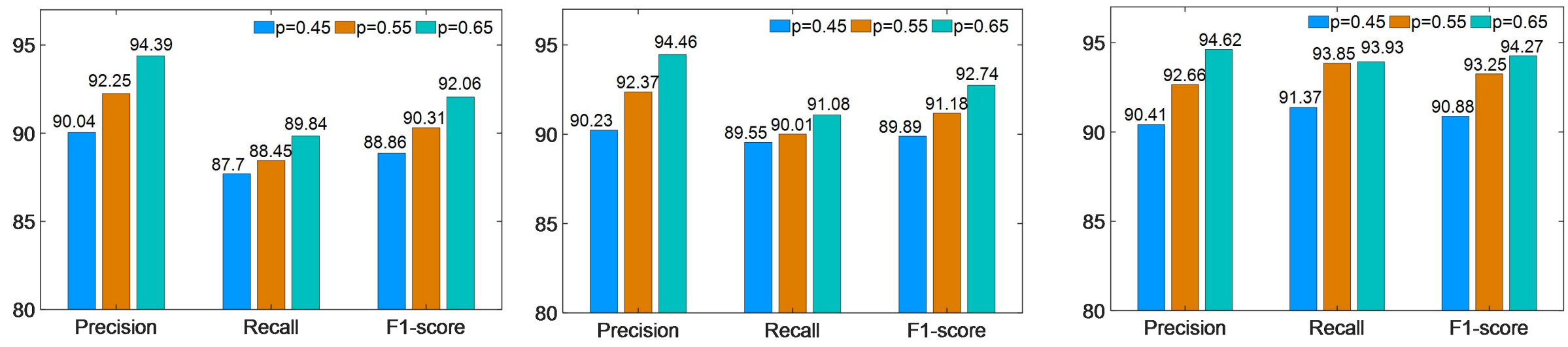

(a) $e$=2 (b) $e$=3 (c) $e$=4

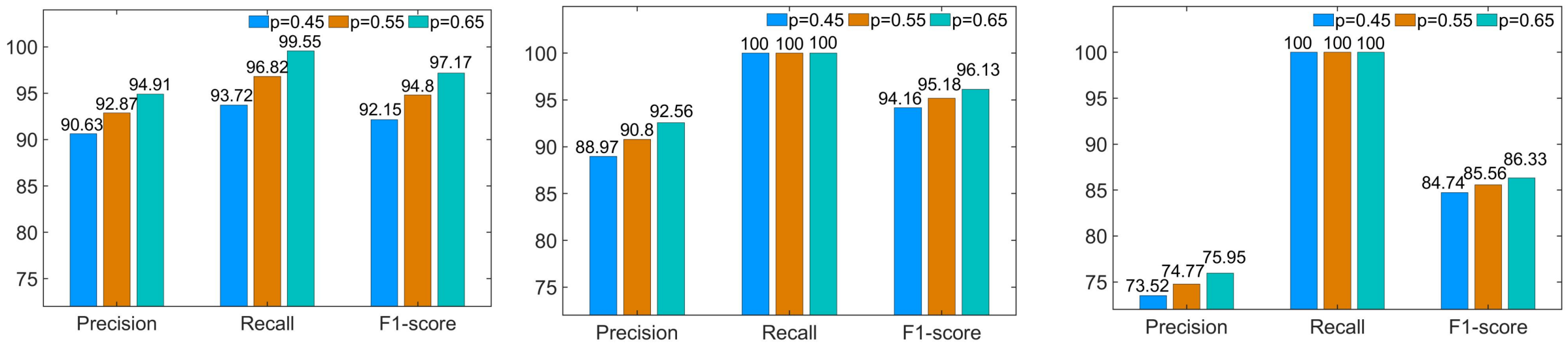

(d) $e$=5 (e) $e$=6 (f) $e$=7

Fig. 7. Comparison of different parameter settings (CMHS)

The comparison results using the 2-classification WADI dataset are shown in Fig. 6(a) - Fig. 6(f). *F1-score* is the parameter that can balance *Precision* and *Recall*. Thus, we consider selecting the parameters that maximize the *F1-score* in this paper. The *Precision* is always higher than 98% with different parameter settings. However, the *Recall* fluctuates from 71.24% to 93%. *F1-score* of 95.96% is the highest when $p$=0.65 if $e$=26 or $e$=27. But the traffic load is smaller when $e$=27, compared with $e$=26. Therefore, we select $p$=0.65 and $e$=27 as the final parameters of our proposed detection approach. In the following content, we refer to our proposed cloud-edge collaborative data anomaly detection approach as G-GCRL.

The comparison results using the 2-classification CMHS dataset are shown in Fig. 7(a) - Fig. 7(f). When $e$=6 or $e$=7, *Recall* is 100% but *F1-score* is not the highest. The reason of *Recall* is 100% is that all normal samples are identified as normal by the sensor data detection model in the edge. However, many abnormal samples are incorrectly classified as normal by this model at the same time, which decreases the *F1-score*. In each subfigure, we find that *F1-score* increases with the increase of $p$, representing that a larger $p$ can make it easier for the extraction of correlations between sensors in CMHS. It is observed that *F1-score* of 97.17% is the highest when $p$=0.65 and $e$=5. Thus, $p$=0.65 and $e$=5 are selected as the parameters of G-GCRL.

#### *4.3.2 Experiment B: Validation of Complex Feature Learning*

Experiment B primarily addresses the following questions: Whether the proposed approach can learn spatial and temporal features of industrial sensor data, by inserting LSTM into GCN?

In Experiment B, the datasets in Table 3 and Table 5 are used for evaluation. The proposed approach is compared with GCN and LSTM to validate the complex feature learning. In Fig. 8(a), the *Recall* and *F1-score* of LSTM are 45.32% and 47.81%, respectively. The *Recall* and *F1-score* of GCN are 75.73% and 85.62%, respectively. However, the *Recall* and *F1-score* of G-GCRL are 92.97% and 95.96% respectively, which are much higher than those of GCN or LSTM. In addition, the *Recall* and *F1-score* of GCN and LSTM are much lower than those of GCRL in Fig. 8(b). This is because LSTM can only extract temporal features and GCN can only extract spatial features. GCRL that is composed of GCN and LSTM can effectively extract the spatial and temporal features of industrial sensor data.

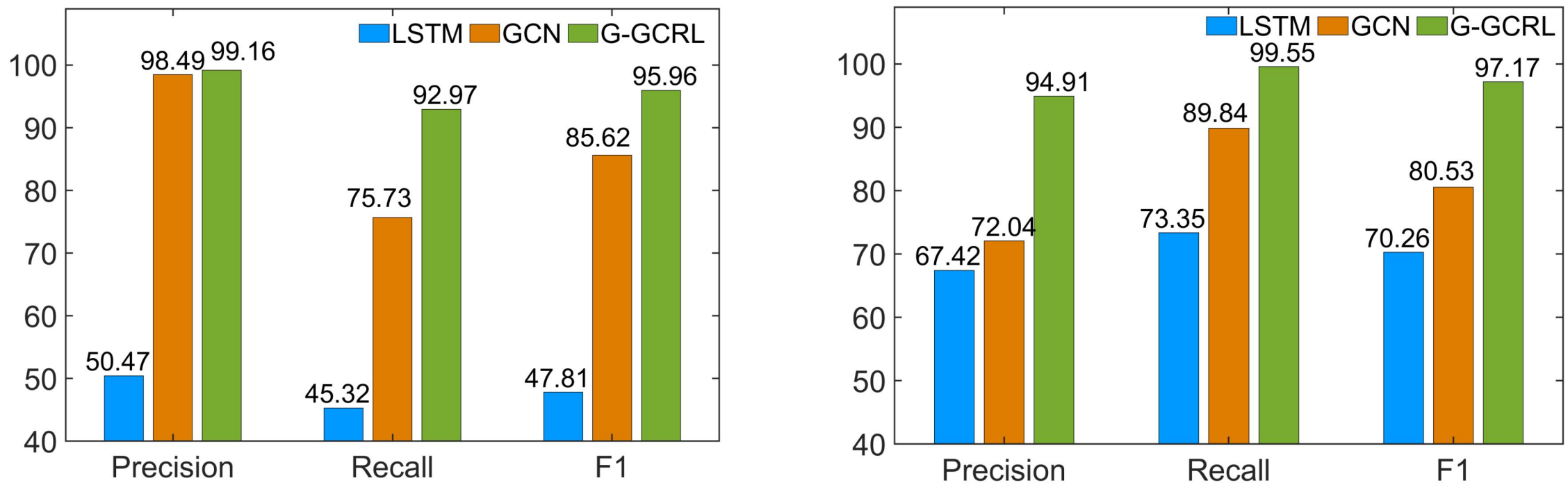

(a) WADI (b) CMHS

Fig. 8. Comparison of different anomaly detection models

### *4.3.3 Experiment C: Compared with Other Sensor Data Anomaly Detection Models*

Experiment C primarily addresses the following questions: How is the performance of our proposed detection approach compared with other sensor data anomaly detection approaches for binary classification?

In Experiment C, the datasets in Table 3 and Table 5 are used for evaluation. In addition, the performance of the proposed G-GCRL is evaluated against 11 different existing models, including AE [29], MAD-GAN [19], LSTM-VAE [62], DAGMM [48], ECAD [55], AQADF [63], GDN [30], GTA [57], B-GAN [64], TCN-GCN [59] and MDAM [65]. These are commonly used baseline models in the field of anomaly detection for industrial sensor data.

AQADF [63]: The Automatic Quasi-periodic time series Anomaly Detection Framework (AQADF) consists of a clustering algorithm and a hybrid LSTM-CNN model with an attention mechanism, which can identify anomalous data.

ECAD [55]: This article proposed an anomaly detection method for industrial sensor data based on cloud-edge collaboration, utilizing neural networks to extract data features for the identification of anomalies.

GTA [57]: Graph learning with Transformer for Anomaly detection (GTA) involves automatically learning a graph structure and graph convolution, and extracts temporal features using a transformer.

B-GAN [64]: Utilizing a GAN structure, the authors leveraged LSTM to augment the number of abnormal samples, effectively resolving the problem of imbalanced distribution between normal and abnormal samples.

TCN-GCN [59]: The author combined 1D-CNN and GCN to analyze the temporal and spatial features. While originally utilized for prediction in the original paper, we can adapt the model to accomplish classification tasks through adjustments.

MDAM [65]: The authors successfully utilize a Multidimensional Deconvolutional network and Attention Mechanism (MDAM) to effectively identify anomalies.

In summary, the models including AE [29], [19], LSTM-VAE [62], DAGMM [48], AQADF [63], B-GAN [64], ECAD [55] and MDAM [65] all employ deep learning algorithms with a focus on extracting temporal features from industrial sensor data. GDN [30] utilizes graph neural networks to focus on extracting spatial features of industrial sensor data. GTA [57] and TCN-GCN [59] are specifically designed to analyze and process both temporal and spatial features of industrial sensor data simultaneously.

Table 8

Performance results of different models

| Dataset | WADI | | | | | CMHS | | | | |
|---|---|---|---|---|---|---|---|---|---|---|
| Model | *Precision* (%) | *Recall* (%) | *F1-score* | *FLOPs* (M) | *MTime* (ms) | *Precision* (%) | *Recall* (%) | *F1-score* | *FLOPs* (M) | *MTime* (ms) |
| AE [29] | 34.35 | 34.35 | 0.34 | **113** | **88** | 38.64 | 40.86 | 0.40 | **102** | **86** |
| MAD-GAN [19] | 41.44 | 33.92 | 0.37 | 168 | 99 | 43.31 | 44.65 | 0.44 | 149 | 94 |
| DAGMM [48] | 54.44 | 26.99 | 0.36 | <u>129</u> | <u>91</u> | 52.52 | 40.86 | 0.46 | <u>114</u> | <u>88</u> |
| LSTM-VAE [62] | 87.79 | 14.45 | 0.25 | 191 | 109 | 66.49 | 51.33 | 0.58 | 175 | 101 |
| AQADF [63] | 89.15 | 40.69 | 0.56 | 226 | 110 | 92.50 | 83.57 | 0.88 | 207 | 106 |
| ECAD [55] | 92.11 | 62.01 | 0.74 | 135 | 92 | 84.89 | 80.64 | 0.83 | 118 | 90 |
| GDN [30] | <u>97.5</u> | 40.19 | 0.57 | 317 | 149 | 84.52 | 77.48 | 0.81 | 295 | 142 |
| GTA [57] | 83.91 | 83.61 | 0.84 | 706 | 307 | 94.10 | 84.91 | 0.89 | 688 | 301 |
| B-GAN [64] | 90.45 | 84.12 | 0.87 | 282 | 127 | 92.13 | 90.22 | 0.91 | 267 | 122 |
| TCN-GCN [59] | 89.27 | 88.19 | 0.87 | 302 | 142 | 93.43 | <u>91.27</u> | 0.92 | 291 | 136 |
| MDAM [65] | 90.59 | **93.66** | <u>0.92</u> | 346 | 161 | <u>94.25</u> | 91.21 | <u>0.93</u> | 320 | 151 |
| G-GCRL | **99.16** | <u>92.97</u> | **0.96** | 390 | 198 | **94.91** | **99.55** | **0.97** | 375 | 183 |

The comparison results of different models are shown in Table 8. The best performance is highlighted in bold and the second-best with underlines. Our proposed data anomaly detection approach based on G-GCRL significantly outperforms all other models for the 2-classification WADI dataset with the performance of *Precision*, *F1-score* up to 99.16%, and 0.96, respectively. In addition, compared to the second-best model, G-GCRL can achieve an overall 4.35% improvement for *F1-score*. For the 2-classification CMHS dataset, our proposed approach can achieve an overall 4.3% improvement for *F1-score*.

The GDN [30] model solely considers spatial features, leading to a relatively low *F1-score*. GTA [57] incorporates GCN to extract spatial features and utilizes transformers for capturing temporal characteristics. However, due to the relatively small scale of the dataset we employed and the complexity of the GTA model structure, overfitting may occur as a consequence of the model's complexity and insufficient training samples, resulting in suboptimal evaluation outcomes. B-GAN [64] addresses the imbalance between positive and negative samples without considering the extraction of temporal-spatial features. TCN-GCN [59] employs 1D-CNN for extracting temporal features and GCN for spatial feature extraction. Nevertheless, the use of convolutional operations in 1D-CNN primarily focuses on capturing local features, where each convolutional kernel is designed to emphasize a specific subset of the data. This characteristic might impede its effectiveness in capturing long-range dependencies, global structural nuances, and temporal relationships intrinsic to sequences. The model [55] employs a cloud-edge computing architecture and uses a lightweight LSTM to extract features from sensor data for anomaly detection tasks. However, the LSTM does not adequately capture the temporal-spatial features of sensor data, particularly neglecting spatial characteristics, resulting in a detection performance that is inferior to the method we propose. The author in [65] employs a MDAM to effectively identify anomalies. However, its computational structure faces challenges in fully extracting spatial features from sensors.

For evaluating the computational complexity and detection latency of the models, this paper adopts two key metrics: FLoating-point OPerations (*FLOPs*) and the Maximum detection time per sample (*Mtime*). *FLOPs* measure the theoretical computational cost of the model during the inference phase and reflect the complexity of the model's computational structure. *Mtime* represents the maximum time required to process a single sample in practical deployment, encompassing data preprocessing, edge-side model detection, data transmission to the cloud, and final analysis by the cloud-side model. As shown in Table 8, AE [29] exhibits the lowest *FLOPs* and the shortest *Mtime* on both the WADI and CMHS datasets, followed by DAGMM [48]. These differences primarily stem from variations in model architecture: AE [29] employs a relatively simple autoencoder for anomaly detection, resulting in lower computational cost and shorter inference time. DAGMM [48] incorporates a Gaussian Mixture Model at the output of the autoencoder. Although this increases the architectural complexity slightly, the overall computational burden remains relatively low. However, despite their advantages in computational cost and inference speed, AE [29] and DAGMM [48] underperform in terms of detection performance, as evidenced by their significantly lower *F1-scores*. This indicates that low computational complexity alone is insufficient to guarantee the practical utility of anomaly detection models in real-world applications. In practice, while *FLOPs* serve as a useful reference for evaluating model efficiency, greater emphasis is placed on the model's responsiveness in actual operating environments—specifically, whether it can meet real-time detection requirements. Generally, if the *Mtime* is shorter than the data sampling interval, the model is considered capable of real-time detection [66]. According to the experimental results, although the proposed G-GCRL incurs relatively higher *FLOPs* and slightly greater computational complexity, its *Mtime* consistently remains below 200 ms, which is well within the 1000 ms sampling interval. Therefore, G-GCRL fully satisfies the real-time detection requirements for practical deployment and demonstrates strong performance in both detection accuracy and system responsiveness.

#### 4.3.4 Experiment D: Validation of Multi-classification Ability

Experiment D primarily addresses the following questions: How is the performance of our proposed detection approach compared with other detection approaches for the classification of multiple anomalies?

In this experiment, the proposed G-GCRL for the classification of multiple anomalies is assessed against AQADF and GCN based on the dataset in Table 3 and Table 5, respectively. The performance metric of detection accuracy for different types of attacks is defined in (16).

$$\mathrm{EACC}_{type_j} = \frac{TN_{type_j}}{number_{type_j}} \tag{16}$$

where $TN_{type_j}$ is the number of correctly classified data in *attack j* dataset. $number_{type_j}$ is the number of data in *attack j* dataset. The comparison results are shown in Fig. 9.

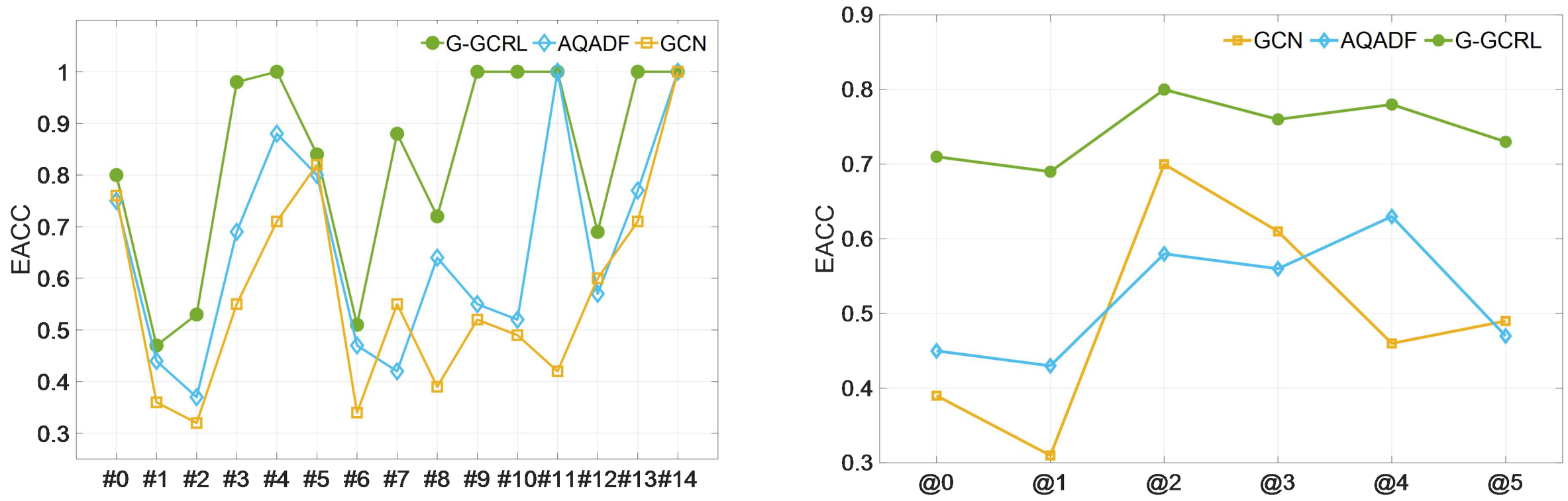

(a) WADI (b) CMHS

Fig. 9. Comparison results of multi-classification

In Fig. 9(a), #0 represents normal and #1 to #14 represent *attack #1* to *attack #14*, respectively. In Fig. 9(b), @5 represents normal and @0 to @4 represents attack @0 to attack@4, respectively. The detection accuracy of each attack of the G-GCRL model is always higher than that of the other models in Fig. 9(a) and Fig. 9(b). This is because the AQADF and GCN model cannot sufficiently extract the spatial and temporal features of industrial sensor data. If the classifier is the G-GCRL model, the accuracy of *attack #4, #9, #10, #11, #13*, and *#14* is almost 100%, respectively. However, the accuracy of *attack #1*, *#2* and *#6* are lower than 0.6 in these three models. After our analysis, we find that these three types of attacks are easily classified as normal. This suggests that the impact of these three attacks on industrial sensor data is relatively small, making them difficult to distinguish. Experiment D validates that our proposed approach outperforms other detection models in multi-class detection performance.

## 5. Conclusions and Future Work

This paper proposes a cloud-edge collaborative data anomaly detection approach for industrial sensor networks. The approach consists of a data preprocess model, a sensor data detection model deployed at individual edges, and a sensor data analysis model deployed in the cloud. The data preprocess model can aggregate collected sensor data and transform it into a Gaussian distribution. The sensor data detection model efficiently filters the substantial volume of sensor data generated during normal operation of the industrial sensor network, and only uploads all the sensor data when the network is in an anomalous state to the cloud. Therefore, this approach can lead to a significant reduction in traffic load. The sensor data analysis model can efficiently extract temporal and spatial features and identify anomalies precisely. Two benchmark datasets WADI and CMHS are used for evaluation compared with other baseline models. The numerical results show that our proposed approach can achieve an impressive 4.35% improvement in *F1-score*, compared with the existing baseline models.

In future work, we will explore adaptive and automated parameter update methods to enhance the model's adaptability to dynamic environments. Furthermore, we will investigate sensor data detection models with higher detection accuracy that can be deployed at the edge to filter out a large volume of normal samples. Moreover, we will explore methodologies to quantify both energy consumption and storage complexity of the algorithms, enabling a more comprehensive evaluation of the performance of our detection models.

## Author Contributions

Conceptualization: Tao Yang.

Software: Tao Yang, Xuefeng Jiang.

Methodology: Wei Li, Jinming Wang.

Visualization: Peiyu Liu, Weijie Hao.

Writing – original draft: Tao Yang.

Writing – review & editing: Qiang Yang.